\documentclass[aps,prl,amsmath,amssymb,superscriptaddress,showpacs,twocolumn]{revtex4-1}

\usepackage{graphicx}
\usepackage{color}
\usepackage{verbatim}

\def\LNO{LaNiO$_3$\,} 
\def\RNO{RNiO$_3$\,}

\begin{document}

\title{Hallmark of strong electronic correlations in LaNiO$_3$:\\ 
photoemission kink and broadening of fully occupied bands}

\author{Xiaoyu Deng}
\affiliation{Centre de Physique Th\'eorique, Ecole Polytechnique, CNRS,
91128 Palaiseau Cedex, France}
\affiliation{Japan Science and Technology Agency, CREST, Kawaguchi
  332-0012, Japan}
\author{Michel Ferrero} 
\affiliation{Centre de Physique Th\'eorique, Ecole Polytechnique, CNRS,
91128 Palaiseau Cedex, France}
\author{Jernej Mravlje}
\affiliation{Centre de Physique Th\'eorique, Ecole Polytechnique, CNRS,
91128 Palaiseau Cedex, France}
\affiliation{Jo\v{z}ef Stefan Institute, Jamova~39, Ljubljana, Slovenia}
\author{Markus Aichhorn}
\affiliation{Institute of Theoretical and Computational Physics, TU Graz, Petersgasse 16, Graz, Austria}
\author{Antoine Georges}
\affiliation{Centre de Physique Th\'eorique, Ecole Polytechnique, CNRS,
91128 Palaiseau Cedex, France}
\affiliation{Coll\`ege de France, 11 place Marcelin Berthelot, 75005 Paris, France}
\affiliation{DPMC, Universit\'e de Gen\`eve, 24 quai Ernest Ansermet, CH-1211 Gen\`eve, Suisse}
\affiliation{Japan Science and Technology Agency, CREST, Kawaguchi
  332-0012, Japan}

\date{\today}

\begin{abstract}
Recent angular-resolved photoemission experiments on \LNO reported 
a renormalization of the Fermi velocity of $e_g$ quasiparticles, a kink in their 
dispersion at $ -0.2$~eV and a large broadening and weakened 
dispersion of the occupied $t_{2g}$ states. 
We show here that all these features result from electronic correlations and 
are quantitatively reproduced by calculations combining density-functional theory 
and dynamical mean-field theory. 
The importance and general relevance of correlation effects in filled bands  
coupled by inter-orbital interactions to a partially-filled band are pointed out. 
\end{abstract}

\maketitle

The effect of strong correlations in partially-filled narrow bands is a central topic 
in condensed matter physics. 
Experimental and theoretical studies have revealed their key role in metal-insulator 
transitions, high-$T_c$ superconductivity and heavy-fermion behavior. 
The remarkable development of sophisticated experimental techniques offers novel
insights, validates established concepts but also  
raises new questions.
Recently, Eguchi \emph{et. al}~\cite{ARPES-Eguchi} performed a delicate 
soft X-ray angular-resolved photoemission (ARPES) study of \LNO, a 
three-dimensional correlated oxide. 
They found clear signatures of strong correlations in the partially filled quasiparticle band  
but also, surprisingly, in completely filled bands.

Among the remarkable family of charge-transfer insulators 
\RNO (with $R$ a rare-earth element) \cite{LNO-phasediagram-Torrance, LNO-struct-Torrance}
, \LNO is the only one which remains metallic down to the lowest temperatures. 
Renewed interest in the nickelates has been triggered by the 
investigation of oxide heterostructures
\cite{stemmer_transport_2011, stemmer_enhancing_2011, hetero-liu-2011,triscone_2011}, and by \LNO being a 
possible electron analogue of cuprate 
superconductors~\cite{LNOtocuprates-Anisimov, turningFS-Hansmann, hetero-Chaloupka-Khaliullin, han_dynamical_2011}. 
The nominal electronic configuration of the nickel orbitals is $t_{2g}^6e_{g}^1$. 
First-principle calculations based on density functional theory (DFT) show that
the $t_{2g}$ bands are fully occupied and that the two nearly-degenerate
$e_g$ bands cross the Fermi level, resulting in a two-sheet Fermi
surface~\cite{DFT-Hamada, balents_RNiO3_prl_2011}. 
Clear experimental evidence for strong electronic correlations, such as enhanced 
effective mass and reduced Drude weight, are revealed by 
the measurement of the low-temperature specific-heat, magnetic susceptibility, 
resistivity and thermopower~\cite{LNO-exp-Rajeev,LNO-exp-Sreedhar, LNO-exp-Xu,LNO-sus-Zhou} 
as well as optical spectroscopy~\cite{LNO-optic-Ouellette,LNO-optic-Stewart,LNO-optic-Stewart-Arxiv}.

The ARPES experiment of Eguchi \emph{et al.} provides important additional insight 
by revealing several interesting features.   
First, there is a renormalization of the $e_g$ quasiparticles Fermi velocity (by a factor of $\sim 1/3$ 
as compared to the DFT band-structure) at the Fermi-surface crossing near the $\Gamma$-point.
Second, the dispersion of these quasiparticles near $\Gamma$ displays a `kink' at a binding energy of order $-0.2$~eV. 
Finally, in the energy range where DFT predicts dispersive $t_{2g}$ bands, the ARPES spectra 
display instead a broad signal which disperses weakly 
and is shifted towards larger binding energy. 
While the Fermi velocity renormalization is consistent with the enhancements observed 
in thermodynamics and transport measurements, these findings raise the following questions.  
(i) What is the physical origin of this renormalization and, especially, of the kink structure? 
(ii) Is the ARPES signal in the $[-1.0,-0.5]$~eV range indeed associated with the $t_{2g}$ bands, 
and if so, why is it broad and weakly dispersing?

In this letter, we address these questions using a combination of 
density functional theory and dynamical mean-field theory (DFT+DMFT), as well as model studies. 
Our results yield a Fermi velocity renormalization and a kink structure in excellent agreement 
with ARPES, which establishes that both  features are a consequence of electronic correlations. 
We also find that the fully filled $t_{2g}$ bands have a weakened dispersion and undergo 
significant broadening.       
This challenges the naive picture 
according to which interactions only have a small effect on fully occupied bands. 
By investigating a simplified model, we explain this finding by the coupling of holes in the  
fully occupied band to particle-hole excitations in the partially occupied one, due to inter-orbital 
interactions. 
This observation is relevant to other correlated materials in which the energy separation between
partially-filled and fully occupied bands is small.

%
%
Our calculations use the full potential implementation of DFT+DMFT detailed 
in Ref.~\onlinecite{LDA+DMFT-Aichhorn}. 
Well-localized Wannier functions \cite{LaFeAsO-Markus} are constructed from a large energy window 
$[-7.8,+3.2]$~eV, which includes all nickel $d$-states and oxygen $p$-states, as 
appropriate for such a charge-transfer compound with significant 
hybridization between those states. 
For simplicity, most calculations were performed for the cubic structure, ignoring 
the small rhomboedral distortion of \LNO.
Local interactions between Ni-$3d$ states are included in the form 
$1/2\,U_{mm'}^{\sigma\sigma'} \hat{n}_{m\sigma}\hat{n}_{m'\sigma'}$. 
We checked that for the problem at hand the omitted spin-flip and pair-hopping terms are not important.
The reduced interaction matrices $U^{\sigma\sigma}_{mm'}$ and $U^{\sigma\bar{\sigma}}_{mm'}$ 
for equal and opposite spins respectively 
are constructed from three Slater integrals $F^0$, $F^2$, $F^4$ in the usual manner. 
We use the standard approximation $F^4/F^2\simeq 0.625$ and define the interaction parameters 
$U$ and $J$ through $U=F^0$, $J=(F^2+F^4)/14$.  
Standard values for nickel compounds~\cite{UJ-Terakura, LNO-optic-Stewart} $U=8.0$~eV and
$J=1.0$~eV are used.  
The double-counting correction term is taken in the `around mean-field'  
form as customary for metallic compounds 
$\Sigma^{\sigma}_{dc}=UN-[U+(M-1)J] N_\sigma/M$, 
where $M$ is the number of correlated orbitals, $N$ is the total occupancy 
and $N_\sigma$ is total occupancy of spin $\sigma$. 
The calculations are fully charge self-consistent, which turns out to be essential in order to insure 
that  the oxygen states are positioned correctly in agreement with PES spectra~\cite{LNO-XPS-Horiba}. 
We solve the DMFT quantum impurity (embedded atom) problem using the
TRIQS~\cite{TRIQS} toolkit and its implementation of the hybridization
expansion continuous-time quantum Monte-Carlo
algorithm~\cite{CTQMC-Werner, CTQMC-Gull-review}, using Legendre
polynomials~\cite{Legendre}. The imaginary-time data is continued
analytically using the stochastic maximum entropy method~\cite{Beach} and Pad\'e
approximants~\cite{Pade}.
All calculations were done at a temperature $T = 1/50$~eV$\simeq 230$~K. 

\begin{figure}[tbp]
  \includegraphics[width=\columnwidth]{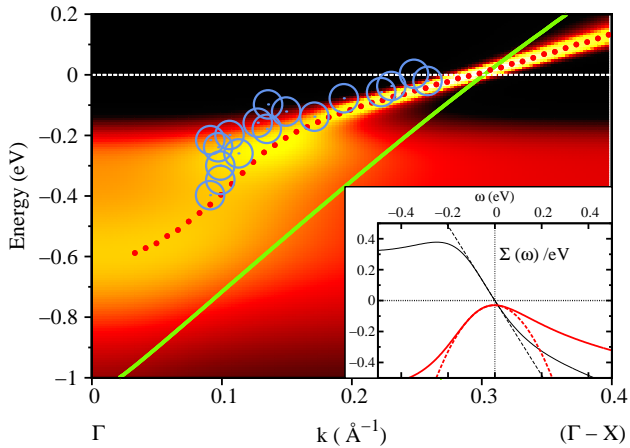}
  \caption{(Color online) Spectral density map close along $\Gamma-X$ (close-up), projected onto $e_g$ character.
Dotted line (red): quasiparticle dispersion setting $\mathrm{Im}\Sigma=0^+$.
Plain line (green): DFT-GGA band.  
Open circles (blue): ARPES data~\cite{ARPES-Eguchi}.  
Inset: Real (black, thin line) and imaginary part (red, thick line) of the 
$e_g$-orbital self-energies.  Dashed lines: linear (parabolic) fit 
of the real (imaginary) part.}
  \label{fig:egspectral}
\end{figure}
We now describe our results. Focusing first on the $e_g$ bands,  
we display on Fig.~\ref{fig:egspectral} the spectral intensity in the
vicinity of the $\Gamma$ point. For clarity, the $t_{2g}$ bands have been
projected out. 
The quasiparticle dispersion obtained by setting the imaginary part of the self-energy 
to $0^+$ is also shown as a guide to the eye (dotted line).
The Fermi velocity is renormalized, corresponding to an effective mass 
$m^*/m\simeq 1/Z \simeq 3$, in good agreement with experiments.  
As one moves further away from the Fermi surface crossing, a `kink'
in the dispersion is found, at an energy close to $-0.2$~eV, 
followed by a ``waterfall'' structure corresponding to a rapid broadening of the spectral function. 
These results agree well with the X-ray ARPES experiments~\cite{ARPES-Eguchi},
providing strong evidence for a purely electronic origin 
of these structures.
Indeed, kinks originating from strong electron correlations have been documented in 
previous DMFT studies at the model level 
and for other materials~\cite{byczuk_kinks_2007, SVO-LDADMFT-Nekrasov,uhrig_kink_prl_2009}. 
The kink can be directly related to the $e_g$ self-energies displayed in
the inset of Fig.~\ref{fig:egspectral}. We note that the low-frequency
Fermi-liquid behavior applies only to a narrow frequency range around 
$\omega=0$. A clear deviation, from linearity for $\mathrm{Re}\Sigma$ and from quadratic behavior for
$\mathrm{Im}\Sigma$, is observed at a negative frequency $\omega\sim-0.2$~eV. 
Below this energy scale, $\mathrm{Re}\Sigma$ has a kink and bends downwards, while 
$\mathrm{Im}\Sigma$ becomes large so that a quasiparticle description no longer applies.

\begin{figure}[tbp]
  \includegraphics[width=\columnwidth]{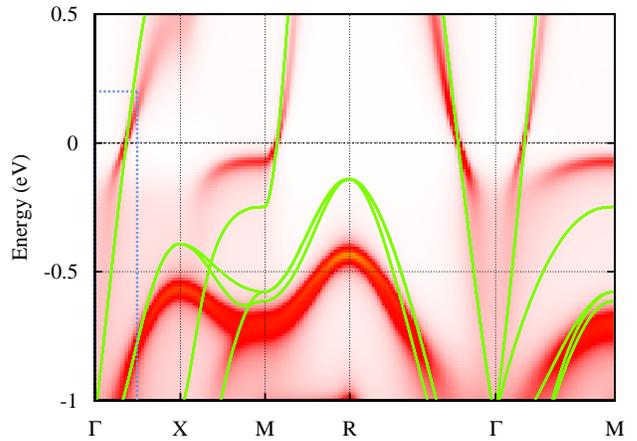}
  \caption{(Color online) Spectral intensity map (all orbitals). 
  DFT-GGA bands are shown for comparison (green, plain lines). The broadening, shifting and weaker dispersion of the $t_{2g}$ states described in the text are clearly apparent. The boundaries of Fig.~\ref{fig:egspectral} are also shown (blue, dotted).}
  \label{fig:Corband_5band}
\end{figure}
Now we turn to the $t_{2g}$ bands.  A full map of the spectral intensity including 
all orbitals is displayed on Fig.~\ref{fig:Corband_5band}, for binding energies 
between $-1$ and $+0.5$~eV.  
It is seen that, in comparison to the DFT-GGA band structure (solid lines), the 
$t_{2g}$ states (i) are pushed further below the Fermi level 
(with e.g. the top of the band at $\sim -0.4$~eV at the $\Gamma$-point 
in contrast to $\sim -0.2$~eV in DFT-GGA ) 
(ii) have a weaker dispersion than the band structure result and  
(iii) importantly, undergo a considerable broadening, due to a rather large scattering rate 
$\mathrm{Im}\Sigma_{t_{2g}}(\omega)$ in the energy range from $\omega\sim -1$~eV to 
$\sim -0.5$~eV, even though the $t_{2g}$ states are fully occupied. 
All three facts agree with the ARPES experiments~\cite{ARPES-Eguchi}, 
which report broad and weakly dispersive spectral weight along 
the $\Gamma$-$X$ and $M$-$R$ directions, and no evidence for dispersive 
$t_{2g}$ bands along $M$-$R$ in the energy range $[-0.5,-0.2]$~eV where DFT-GGA 
would place these bands. 
A refined quantitative positioning of the $t_{2g}$ states would require to take the 
rhomboedral distortion of \LNO into account. 

\begin{figure}[tbp]
  \includegraphics[width=0.5\columnwidth]{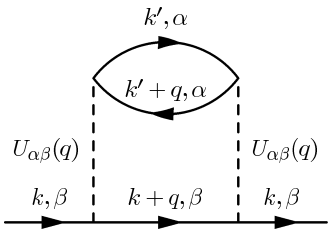}
  \caption{Second-order contribution to the self-energy.}
  \label{fig:diag}
\end{figure}
Hence, it is seen that the effects of interactions on one-particle spectra are strong 
for the $t_{2g}$ bands, in spite of those bands being completely filled. 
This may seem surprising if one follows naive intuition drawn 
from single-band models. 
However, in a multi-band context this is quite natural: a hole created in the 
occupied $t_{2g}$ band (as in a photoemission process) can scatter against 
particle-hole excitations associated with the partially occupied $e_g$ states, 
due to the {\it inter-orbital} matrix elements.
Consider for example the second-order diagram for the $t_{2g}$ self-energy 
shown in Fig.~\ref{fig:diag} in which a hole in an occupied band $\beta$ couples through 
$U_{\alpha\beta}(q)$ to 
a particle-hole excitation of band $\alpha$ described by the corresponding  polarization bubble. 
The latter is proportional to 
$n_{\alpha,k'+q}-n_{\alpha,k'}$
 and would vanish for a 
fully filled or empty band, but gives a non-zero contribution if $\alpha$ is partially occupied. 
This conclusion
persists for higher-order diagrams. 
This effect has similarities with the physics of X-ray edge singularities and core-hole photoemission, 
with the important difference that the hole is not static in the present case.

In order to understand this important effect in a simpler setting, 
we consider a two-band Hubbard model  
$H=\sum_{k,m\sigma}\epsilon_{k}c^{\dag}_{km\sigma}c_{km\sigma}-\Delta \sum_i\hat{n}_{i,2\sigma} 
+U\sum_{i}\sum_{i,m\sigma<m'\sigma'}\hat{n}_{i,m\sigma}\hat{n}_{i,m'\sigma'}$. 
In this expression, $m=1,2$ is a band index and $\Delta$ a crystal-field splitting 
such that the second band is lower in energy. 
We take both bands to have a semicircular density of states with the same half-bandwidth 
$D$, $U/D=2.0$, and no inter-band hybridization. 
We set the chemical potential so that there are 3 electrons per site. 
\begin{figure}[tbp]
  \includegraphics[width=\columnwidth]{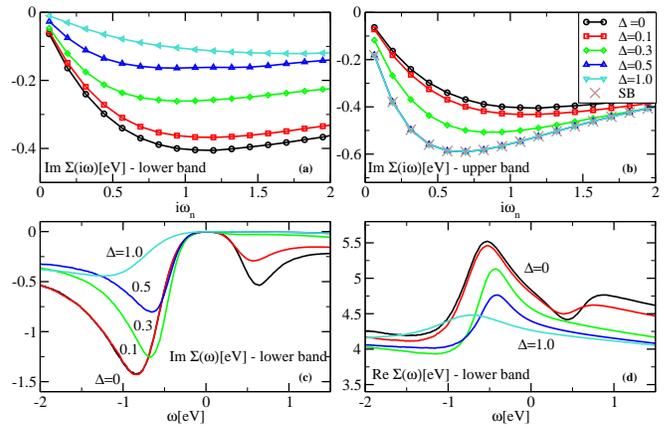}
  \caption{(Color online) Self-energies of the two-band model with increasing 
    crystal-field splitting $\Delta$. 
    (a-b): Matsubara-frequency self-energy $\mathrm{Im}\Sigma_m(i\omega_n)$ 
    of (a) the lower band (fully filled for $\Delta\geq 0.5D $), 
    (b) the higher band (half-filled for $\Delta\geq 0.5D$). Single-band (SB) results are shown with crosses.
    (c-d): Imaginary (c) and real (d) part of the real-frequency self-energy for
    the lower band (analytic continuation using stochastic maximum entropy).}
  \label{fig:sigmamodel}
\end{figure}
This model captures the transition from a two-band metal into a
single-band metal as the crystal-field splitting is progressively 
increased~\cite{Model-Poteryaev}. 

On Fig.~\ref{fig:sigmamodel}(a)(b) we display 
$\mathrm{Im}\Sigma_m(i\omega_n)$ for imaginary (Matsubara) frequencies, 
for the lower and upper bands, respectively.  
The strength of correlations is characterized by the overall magnitude of 
$\Sigma_m(i\omega_n)$. 
As $\Delta$ increases, the occupancy of the upper band progressively diminishes 
from $n_1=1.5$ ($\Delta=0$) to $n_1=1$ and the strength of correlations 
in the upper band increases, reflected also in
$Z^{-1}=1-\partial \Sigma(i\omega)/\partial(i\omega)|_{\omega \to 0}$ 
corresponding to the enhancement of effective mass.
As soon as the lower band is completely filled ($n_2=2$), which happens here at $\Delta=0.5 D$,
$\Sigma_1(i\omega_n)$ saturates and becomes identical to the self-energy obtained in
the single-band case $\Delta \to \infty$ (Fig.~\ref{fig:sigmamodel}b).
This is in agreement with the qualitative arguments presented above: 
no particle-hole excitations can be created (at low energy) in the fully occupied band. 

Correspondingly, the correlations in the lower band progressively diminish, as the 
band is filled-in (Fig.~\ref{fig:sigmamodel}a). However, in contrast to the upper band, 
no qualitative change is seen at $\Delta=0.5 D$ and the correlations evolve 
smoothly as $\Delta$ is further increased, remaining sizable even at $\Delta=1.0 D$. 
On Figs.~\ref{fig:sigmamodel}(c)-(d) we display $\mathrm{Im}\Sigma_2(\omega+i0^+)$ 
and $\mathrm{Re}\Sigma_2(\omega+i0^+)$ analytically continued on the real-frequency 
axis. At $\Delta=0$, the self energy has a pronounced peak at negative frequencies, 
corresponding to the pronounced lower Hubbard band. This structure  
persists through the regime $\Delta \gtrsim 0.5$ when the lower band is totally filled. 
\begin{figure}[tbp]
  \includegraphics[width=\columnwidth]{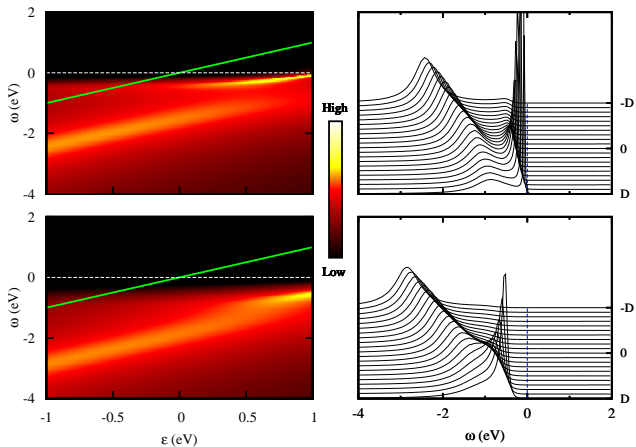}
  \caption{(Color online) Energy-resolved spectral function for the two-band model 
  $A_2(\epsilon,\omega)=
-\mathrm{Im}\left[\omega+\mu-\epsilon-\Delta-\Sigma_2(\omega+i0^+)\right]^{-1}/\pi$. 
   Left: contour map in the $(\epsilon,\omega)$ plane, with $-D\leq\epsilon\leq+D$. 
   Right: energy ($\omega$)-resolved spectra for different energies $\epsilon+\Delta$ 
   in the band. 
   Upper panel: $\Delta=0.5D$. Lower panel: $\Delta=1.0D$.}
  \label{fig:Aew_model}
\end{figure}
To complete the picture, we display on Fig.~\ref{fig:Aew_model} 
the energy-resolved spectral functions of the lower band.  
At the transition point $\Delta=0.5 D$, the top of the lower band touches the 
Fermi level and develops there a sharp quasi-particle peak at $\omega=0$. 
Most of the spectral weight is carried, however, by a broad structure seen at 
negative frequencies.
When the lower band lies below the Fermi level ($\Delta=1.0 D$),  
the  quasiparticle structure merges with the broad Hubbard-like and 
the energy-resolved spectra are broad for most values of the momentum 
(except at the very top of the band). 
Hence, these model results fully confirm the qualitative expectations above and 
clearly demonstrate strong correlation effects in the fully occupied band due to 
inter-orbital interactions. 

%
In conclusion, we have shown that electronic correlations explain the spectral features 
seen in photoemission experiments on \LNO. 
Our DFT+DMFT results are in quantitative agreement with these measurements.
While a description in terms of renormalized quasiparticles applies 
at low-energy~\cite{balents_MIT_arXiv}, the `kink' in the quasiparticle dispersion, 
shown here to be of electronic origin, signals the breakdown of this picture. 
We have demonstrated that inter-orbital interactions are responsible
for the observed large broadening of the fully occupied $t_{2g}$
states. These correlation effects become important when the occupied
band is close to the Fermi energy and hence are relevant to other materials, 
for example to the broadening of $e'_g$ states in
Na$_x$CoO$_2$~\cite{hasan_prl_2004, yang_prl_2005, nicolaou_prl_2010}.
On a broader level, our results question the applicability of band theory to occupied 
bands strongly coupled to partially-filled ones.  
%

We are most grateful to L.~Balents for discussions which triggered this project. 
We also acknowledge discussions with S.~Biermann, with  R.~Eguchi, K.~Horiba and
S.~Shin about their ARPES measurements, 
with R.~Scherwitzl, S.~Gariglio, P.~Zubko, M.~Gibert and J.-M.~Triscone about transport in nickelate heterostructures. 
TRIQS uses some libraries of the ALPS~\cite{ALPS} project.  
Support was provided by the Partner University Fund (PUF), ICAM and
the Swiss National Foundation MaNEP program. M.A. acknowledges
financial support from the Austrian Science Fund, project F4103, and
hospitality at Ecole Polytechnique.


\begin{thebibliography}{10}%
\makeatletter
\providecommand \@ifxundefined [1]{%
 \ifx #1\undefined \expandafter \@firstoftwo
 \else \expandafter \@secondoftwo
\fi
}%
\providecommand \@ifnum [1]{%
 \ifnum #1\expandafter \@firstoftwo
 \else \expandafter \@secondoftwo
\fi
}%
\providecommand \enquote [1]{``#1''}%
\providecommand \bibnamefont  [1]{#1}%
\providecommand \bibfnamefont [1]{#1}%
\providecommand \citenamefont [1]{#1}%
\providecommand\href[0]{\@sanitize\@href}%
\providecommand\@href[1]{\endgroup\@@startlink{#1}\endgroup\@@href}%
\providecommand\@@href[1]{#1\@@endlink}%
\providecommand \@sanitize [0]{\begingroup\catcode`\&12\catcode`\#12\relax}%
\@ifxundefined \pdfoutput {\@firstoftwo}{%
 \@ifnum{\z@=\pdfoutput}{\@firstoftwo}{\@secondoftwo}%
}{%
 \providecommand\@@startlink[1]{\leavevmode}%
 \providecommand\@@endlink[0]{}%
}{%
 \providecommand\@@startlink[1]{%
  \leavevmode
  \pdfstartlink
   attr{/Border[0 0 1 ]/H/I/C[0 1 1]}%
   user{/Subtype/Link/A<</Type/Action/S/URI/URI(#1)>>}%
  \relax
 }%
 \providecommand\@@endlink[0]{\pdfendlink}%
}%
\providecommand \url  [0]{\begingroup\@sanitize \@url }%
\providecommand \@url [1]{\endgroup\@href {#1}{\urlprefix}}%
\providecommand \urlprefix [0]{URL }%
\providecommand \Eprint[0]{\href }%
\@ifxundefined \urlstyle {%
  \providecommand \doi [1]{doi:\discretionary{}{}{}#1}%
}{%
  \providecommand \doi [0]{doi:\discretionary{}{}{}\begingroup
  \urlstyle{rm}\Url }%
}%
\providecommand \doibase [0]{http://dx.doi.org/}%
\providecommand \Doi[1]{\href{\doibase#1}}%
\providecommand \bibAnnote [3]{%
  \BibitemShut{#1}%
  \begin{quotation}\noindent
    \textsc{Key:}\ #2\\\textsc{Annotation:}\ #3%
  \end{quotation}%
}%
\providecommand \bibAnnoteFile [2]{%
  \IfFileExists{#2}{\bibAnnote {#1} {#2} {\input{#2}}}{}%
}%
\providecommand \typeout [0]{\immediate \write \m@ne }%
\providecommand \selectlanguage [0]{\@gobble}%
\providecommand \bibinfo [0]{\@secondoftwo}%
\providecommand \bibfield [0]{\@secondoftwo}%
\providecommand \translation [1]{[#1]}%
\providecommand \BibitemOpen[0]{}%
\providecommand \bibitemStop [0]{}%
\providecommand \bibitemNoStop [0]{.\EOS\space}%
\providecommand \EOS [0]{\spacefactor3000\relax}%
\providecommand \BibitemShut [1]{\csname bibitem#1\endcsname}%
\bibitem{ARPES-Eguchi}%
  \BibitemOpen
  \bibfield{author}{%
  \bibinfo {author} {\bibfnamefont{R.}~\bibnamefont{Eguchi}}, \bibinfo {author}
  {\bibfnamefont{A.}~\bibnamefont{Chainani}}, \bibinfo {author}
  {\bibfnamefont{M.}~\bibnamefont{Taguchi}}, \bibinfo {author}
  {\bibfnamefont{M.}~\bibnamefont{Matsunami}}, \bibinfo {author}
  {\bibfnamefont{Y.}~\bibnamefont{Ishida}}, \bibinfo {author}
  {\bibfnamefont{K.}~\bibnamefont{Horiba}}, \bibinfo {author}
  {\bibfnamefont{Y.}~\bibnamefont{Senba}}, \bibinfo {author}
  {\bibfnamefont{H.}~\bibnamefont{Ohashi}},\ and\ \bibinfo {author}
  {\bibfnamefont{S.}~\bibnamefont{Shin}},\ }%
  \bibfield{journal}{%
  \Doi{10.1103/PhysRevB.79.115122}{\bibinfo {journal} {Phys. Rev. B}}\ }%
  \textbf{\bibinfo {volume} {79}},\ \bibinfo {pages} {115122} (\bibinfo {year}
  {2009})%
  \bibAnnoteFile{NoStop}{ARPES-Eguchi}%
\bibitem{LNO-phasediagram-Torrance}%
  \BibitemOpen
  \bibfield{author}{%
  \bibinfo {author} {\bibfnamefont{J.~B.}\ \bibnamefont{Torrance}}, \bibinfo
  {author} {\bibfnamefont{P.}~\bibnamefont{Lacorre}}, \bibinfo {author}
  {\bibfnamefont{A.~I.}\ \bibnamefont{Nazzal}}, \bibinfo {author}
  {\bibfnamefont{E.~J.}\ \bibnamefont{Ansaldo}},\ and\ \bibinfo {author}
  {\bibfnamefont{C.}~\bibnamefont{Niedermayer}},\ }%
  \bibfield{journal}{%
  \Doi{10.1103/PhysRevB.45.8209}{\bibinfo {journal} {Phys. Rev. B}}\ }%
  \textbf{\bibinfo {volume} {45}},\ \bibinfo {pages} {8209} (\bibinfo {year}
  {1992})%
  \bibAnnoteFile{NoStop}{LNO-phasediagram-Torrance}%
\bibitem{LNO-struct-Torrance}%
  \BibitemOpen
  \bibfield{author}{%
  \bibinfo {author} {\bibfnamefont{J.~L.}\ \bibnamefont{Garc\'\i{}a-Mu\~noz}},
  \bibinfo {author} {\bibfnamefont{J.}~\bibnamefont{Rodr\'\i{}guez-Carvajal}},
  \bibinfo {author} {\bibfnamefont{P.}~\bibnamefont{Lacorre}},\ and\ \bibinfo
  {author} {\bibfnamefont{J.~B.}\ \bibnamefont{Torrance}},\ }%
  \bibfield{journal}{%
  \Doi{10.1103/PhysRevB.46.4414}{\bibinfo {journal} {Phys. Rev. B}}\ }%
  \textbf{\bibinfo {volume} {46}},\ \bibinfo {pages} {4414} (\bibinfo {year}
  {1992})%
  \bibAnnoteFile{NoStop}{LNO-struct-Torrance}%
\bibitem{stemmer_transport_2011}%
  \BibitemOpen
  \bibfield{author}{%
  \bibinfo {author} {\bibfnamefont{P.}~\bibnamefont{Moetakef}}, \bibinfo
  {author} {\bibfnamefont{J.~Y.}\ \bibnamefont{Zhang}}, \bibinfo {author}
  {\bibfnamefont{A.}~\bibnamefont{Kozhanov}}, \bibinfo {author}
  {\bibfnamefont{B.}~\bibnamefont{Jalan}}, \bibinfo {author}
  {\bibfnamefont{R.}~\bibnamefont{Seshadri}}, \bibinfo {author}
  {\bibfnamefont{S.~J.}\ \bibnamefont{Allen}},\ and\ \bibinfo {author}
  {\bibfnamefont{S.}~\bibnamefont{Stemmer}},\ }%
  \bibfield{journal}{%
  \Doi{10.1063/1.3568894}{\bibinfo {journal} {Appl. Phys. Lett.}}\ }%
  \textbf{\bibinfo {volume} {98}},\ \bibinfo {pages} {112110} (\bibinfo {year}
  {2011})%
  \bibAnnoteFile{NoStop}{stemmer_transport_2011}%
\bibitem{stemmer_enhancing_2011}%
  \BibitemOpen
  \bibfield{author}{%
  \bibinfo {author} {\bibfnamefont{B.}~\bibnamefont{Jalan}}, \bibinfo {author}
  {\bibfnamefont{S.~J.}\ \bibnamefont{Allen}}, \bibinfo {author}
  {\bibfnamefont{G.~E.}\ \bibnamefont{Beltz}}, \bibinfo {author}
  {\bibfnamefont{P.}~\bibnamefont{Moetakef}},\ and\ \bibinfo {author}
  {\bibfnamefont{S.}~\bibnamefont{Stemmer}},\ }%
  \bibfield{journal}{%
  \Doi{10.1063/1.3571447}{\bibinfo {journal} {Appl. Phys. Lett.}}\ }%
  \textbf{\bibinfo {volume} {98}},\ \bibinfo {pages} {132102} (\bibinfo {year}
  {2011})%
  \bibAnnoteFile{NoStop}{stemmer_enhancing_2011}%
\bibitem{hetero-liu-2011}%
  \BibitemOpen
  \bibfield{author}{%
  \bibinfo {author} {\bibfnamefont{J.}~\bibnamefont{Liu}}, \bibinfo {author}
  {\bibfnamefont{S.}~\bibnamefont{Okamoto}}, \bibinfo {author}
  {\bibfnamefont{M.}~\bibnamefont{van Veenendaal}}, \bibinfo {author}
  {\bibfnamefont{M.}~\bibnamefont{Kareev}}, \bibinfo {author}
  {\bibfnamefont{B.}~\bibnamefont{Gray}}, \bibinfo {author}
  {\bibfnamefont{P.}~\bibnamefont{Ryan}}, \bibinfo {author}
  {\bibfnamefont{J.~W.}\ \bibnamefont{Freeland}},\ and\ \bibinfo {author}
  {\bibfnamefont{J.}~\bibnamefont{Chakhalian}},\ }%
  \bibfield{journal}{%
  \Doi{10.1103/PhysRevB.83.161102}{\bibinfo {journal} {Phys. Rev. B}}\ }%
  \textbf{\bibinfo {volume} {83}},\ \bibinfo {pages} {161102} (\bibinfo {year}
  {2011})%
  \bibAnnoteFile{NoStop}{hetero-liu-2011}%
\bibitem{triscone_2011}%
  \BibitemOpen
  \bibfield{author}{%
  \bibinfo {author} {\bibfnamefont{R.}~\bibnamefont{Scherwitzl}}, \bibinfo
  {author} {\bibfnamefont{S.}~\bibnamefont{Gariglio}}, \bibinfo {author}
  {\bibfnamefont{M.}~\bibnamefont{Gabay}}, \bibinfo {author}
  {\bibfnamefont{P.}~\bibnamefont{Zubko}}, \bibinfo {author}
  {\bibfnamefont{M.}~\bibnamefont{Gibert}},\ and\ \bibinfo {author}
  {\bibfnamefont{J.-M.}\ \bibnamefont{Triscone}},\ }%
  \bibfield{journal}{%
  \Doi{10.1103/PhysRevLett.106.246403}{\bibinfo {journal} {Phys. Rev. Lett.}}\
  }%
  \textbf{\bibinfo {volume} {106}},\ \bibinfo {pages} {246403} (\bibinfo {year}
  {2011})%
  \bibAnnoteFile{NoStop}{triscone_2011}%
\bibitem{LNOtocuprates-Anisimov}%
  \BibitemOpen
  \bibfield{author}{%
  \bibinfo {author} {\bibfnamefont{V.~I.}\ \bibnamefont{Anisimov}}, \bibinfo
  {author} {\bibfnamefont{D.}~\bibnamefont{Bukhvalov}},\ and\ \bibinfo {author}
  {\bibfnamefont{T.~M.}\ \bibnamefont{Rice}},\ }%
  \bibfield{journal}{%
  \Doi{10.1103/PhysRevB.59.7901}{\bibinfo {journal} {Phys. Rev. B}}\ }%
  \textbf{\bibinfo {volume} {59}},\ \bibinfo {pages} {7901} (\bibinfo {year}
  {1999})%
  \bibAnnoteFile{NoStop}{LNOtocuprates-Anisimov}%
\bibitem{turningFS-Hansmann}%
  \BibitemOpen
  \bibfield{author}{%
  \bibinfo {author} {\bibfnamefont{P.}~\bibnamefont{Hansmann}}, \bibinfo
  {author} {\bibfnamefont{X.}~\bibnamefont{Yang}}, \bibinfo {author}
  {\bibfnamefont{A.}~\bibnamefont{Toschi}}, \bibinfo {author}
  {\bibfnamefont{G.}~\bibnamefont{Khaliullin}}, \bibinfo {author}
  {\bibfnamefont{O.~K.}\ \bibnamefont{Andersen}},\ and\ \bibinfo {author}
  {\bibfnamefont{K.}~\bibnamefont{Held}},\ }%
  \bibfield{journal}{%
  \Doi{10.1103/PhysRevLett.103.016401}{\bibinfo {journal} {Phys. Rev. Lett.}}\
  }%
  \textbf{\bibinfo {volume} {103}},\ \bibinfo {pages} {016401} (\bibinfo {year}
  {2009})%
  \bibAnnoteFile{NoStop}{turningFS-Hansmann}%
\bibitem{hetero-Chaloupka-Khaliullin}%
  \BibitemOpen
  \bibfield{author}{%
  \bibinfo {author} {\bibfnamefont{J.}~\bibnamefont{Chaloupka}}\ and\ \bibinfo
  {author} {\bibfnamefont{G.}~\bibnamefont{Khaliullin}},\ }%
  \bibfield{journal}{%
  \Doi{10.1103/PhysRevLett.100.016404}{\bibinfo {journal} {Phys. Rev. Lett.}}\
  }%
  \textbf{\bibinfo {volume} {100}},\ \bibinfo {pages} {016404} (\bibinfo {year}
  {2008})%
  \bibAnnoteFile{NoStop}{hetero-Chaloupka-Khaliullin}%
\bibitem{han_dynamical_2011}%
  \BibitemOpen
  \bibfield{author}{%
  \bibinfo {author} {\bibfnamefont{M.~J.}\ \bibnamefont{Han}}, \bibinfo
  {author} {\bibfnamefont{X.}~\bibnamefont{Wang}}, \bibinfo {author}
  {\bibfnamefont{C.~A.}\ \bibnamefont{Marianetti}},\ and\ \bibinfo {author}
  {\bibfnamefont{A.~J.}\ \bibnamefont{Millis}},\ }%
  \Eprint{http://arxiv.org/abs/1105.0016}{arXiv:1105.0016}%
  \bibAnnoteFile{NoStop}{han_dynamical_2011}%
\bibitem{DFT-Hamada}%
  \BibitemOpen
  \bibfield{author}{%
  \bibinfo {author} {\bibfnamefont{N.}~\bibnamefont{Hamada}},\ }%
  \bibfield{journal}{%
  \Doi{DOI: 10.1016/0022-3697(93)90159-O}{\bibinfo {journal} {J. Phys. Chem.
  Solids}}\ }%
  \textbf{\bibinfo {volume} {54}},\ \bibinfo {pages} {1157 } (\bibinfo {year}
  {1993})%
  \bibAnnoteFile{NoStop}{DFT-Hamada}%
\bibitem{balents_RNiO3_prl_2011}%
  \BibitemOpen
  \bibfield{author}{%
  \bibinfo {author} {\bibfnamefont{S.}~\bibnamefont{Lee}}, \bibinfo {author}
  {\bibfnamefont{R.}~\bibnamefont{Chen}},\ and\ \bibinfo {author}
  {\bibfnamefont{L.}~\bibnamefont{Balents}},\ }%
  \bibfield{journal}{%
  \Doi{10.1103/PhysRevLett.106.016405}{\bibinfo {journal} {Phys. Rev. Lett.}}\
  }%
  \textbf{\bibinfo {volume} {106}},\ \bibinfo {pages} {016405} (\bibinfo {year}
  {2011})%
  \bibAnnoteFile{NoStop}{balents_RNiO3_prl_2011}%
\bibitem{LNO-exp-Rajeev}%
  \BibitemOpen
  \bibfield{author}{%
  \bibinfo {author} {\bibfnamefont{K.}~\bibnamefont{Rajeev}}, \bibinfo {author}
  {\bibfnamefont{G.}~\bibnamefont{Shivashankar}},\ and\ \bibinfo {author}
  {\bibfnamefont{A.}~\bibnamefont{Raychaudhuri}},\ }%
  \bibfield{journal}{%
  \Doi{DOI: 10.1016/0038-1098(91)90915-I}{\bibinfo {journal} {Solid State
  Commun.}}\ }%
  \textbf{\bibinfo {volume} {79}},\ \bibinfo {pages} {591 } (\bibinfo {year}
  {1991})%
  \bibAnnoteFile{NoStop}{LNO-exp-Rajeev}%
\bibitem{LNO-exp-Sreedhar}%
  \BibitemOpen
  \bibfield{author}{%
  \bibinfo {author} {\bibfnamefont{K.}~\bibnamefont{Sreedhar}}, \bibinfo
  {author} {\bibfnamefont{J.~M.}\ \bibnamefont{Honig}}, \bibinfo {author}
  {\bibfnamefont{M.}~\bibnamefont{Darwin}}, \bibinfo {author}
  {\bibfnamefont{M.}~\bibnamefont{McElfresh}}, \bibinfo {author}
  {\bibfnamefont{P.~M.}\ \bibnamefont{Shand}}, \bibinfo {author}
  {\bibfnamefont{J.}~\bibnamefont{Xu}}, \bibinfo {author}
  {\bibfnamefont{B.~C.}\ \bibnamefont{Crooker}},\ and\ \bibinfo {author}
  {\bibfnamefont{J.}~\bibnamefont{Spalek}},\ }%
  \bibfield{journal}{%
  \Doi{10.1103/PhysRevB.46.6382}{\bibinfo {journal} {Phys. Rev. B}}\ }%
  \textbf{\bibinfo {volume} {46}},\ \bibinfo {pages} {6382} (\bibinfo {year}
  {1992})%
  \bibAnnoteFile{NoStop}{LNO-exp-Sreedhar}%
\bibitem{LNO-exp-Xu}%
  \BibitemOpen
  \bibfield{author}{%
  \bibinfo {author} {\bibfnamefont{X.~Q.}\ \bibnamefont{Xu}}, \bibinfo {author}
  {\bibfnamefont{J.~L.}\ \bibnamefont{Peng}}, \bibinfo {author}
  {\bibfnamefont{Z.~Y.}\ \bibnamefont{Li}}, \bibinfo {author}
  {\bibfnamefont{H.~L.}\ \bibnamefont{Ju}},\ and\ \bibinfo {author}
  {\bibfnamefont{R.~L.}\ \bibnamefont{Greene}},\ }%
  \bibfield{journal}{%
  \Doi{10.1103/PhysRevB.48.1112}{\bibinfo {journal} {Phys. Rev. B}}\ }%
  \textbf{\bibinfo {volume} {48}},\ \bibinfo {pages} {1112} (\bibinfo {year}
  {1993})%
  \bibAnnoteFile{NoStop}{LNO-exp-Xu}%
\bibitem{LNO-sus-Zhou}%
  \BibitemOpen
  \bibfield{author}{%
  \bibinfo {author} {\bibfnamefont{J.-S.}\ \bibnamefont{Zhou}}, \bibinfo
  {author} {\bibfnamefont{J.~B.}\ \bibnamefont{Goodenough}}, \bibinfo {author}
  {\bibfnamefont{B.}~\bibnamefont{Dabrowski}}, \bibinfo {author}
  {\bibfnamefont{P.~W.}\ \bibnamefont{Klamut}},\ and\ \bibinfo {author}
  {\bibfnamefont{Z.}~\bibnamefont{Bukowski}},\ }%
  \bibfield{journal}{%
  \Doi{10.1103/PhysRevLett.84.526}{\bibinfo {journal} {Phys. Rev. Lett.}}\ }%
  \textbf{\bibinfo {volume} {84}},\ \bibinfo {pages} {526} (\bibinfo {year}
  {2000})%
  \bibAnnoteFile{NoStop}{LNO-sus-Zhou}%
\bibitem{LNO-optic-Ouellette}%
  \BibitemOpen
  \bibfield{author}{%
  \bibinfo {author} {\bibfnamefont{D.~G.}\ \bibnamefont{Ouellette}}, \bibinfo
  {author} {\bibfnamefont{S.}~\bibnamefont{Lee}}, \bibinfo {author}
  {\bibfnamefont{J.}~\bibnamefont{Son}}, \bibinfo {author}
  {\bibfnamefont{S.}~\bibnamefont{Stemmer}}, \bibinfo {author}
  {\bibfnamefont{L.}~\bibnamefont{Balents}}, \bibinfo {author}
  {\bibfnamefont{A.~J.}\ \bibnamefont{Millis}},\ and\ \bibinfo {author}
  {\bibfnamefont{S.~J.}\ \bibnamefont{Allen}},\ }%
  \bibfield{journal}{%
  \Doi{10.1103/PhysRevB.82.165112}{\bibinfo {journal} {Phys. Rev. B}}\ }%
  \textbf{\bibinfo {volume} {82}},\ \bibinfo {pages} {165112} (\bibinfo {year}
  {2010})%
  \bibAnnoteFile{NoStop}{LNO-optic-Ouellette}%
\bibitem{LNO-optic-Stewart}%
  \BibitemOpen
  \bibfield{author}{%
  \bibinfo {author} {\bibfnamefont{M.~K.}\ \bibnamefont{Stewart}}, \bibinfo
  {author} {\bibfnamefont{C.-H.}\ \bibnamefont{Yee}}, \bibinfo {author}
  {\bibfnamefont{J.}~\bibnamefont{Liu}}, \bibinfo {author}
  {\bibfnamefont{M.}~\bibnamefont{Kareev}}, \bibinfo {author}
  {\bibfnamefont{R.~K.}\ \bibnamefont{Smith}}, \bibinfo {author}
  {\bibfnamefont{B.~C.}\ \bibnamefont{Chapler}}, \bibinfo {author}
  {\bibfnamefont{M.}~\bibnamefont{Varela}}, \bibinfo {author}
  {\bibfnamefont{P.~J.}\ \bibnamefont{Ryan}}, \bibinfo {author}
  {\bibfnamefont{K.}~\bibnamefont{Haule}}, \bibinfo {author}
  {\bibfnamefont{J.}~\bibnamefont{Chakhalian}},\ and\ \bibinfo {author}
  {\bibfnamefont{D.~N.}\ \bibnamefont{Basov}},\ }%
  \bibfield{journal}{%
  \Doi{10.1103/PhysRevB.83.075125}{\bibinfo {journal} {Phys. Rev. B}}\ }%
  \textbf{\bibinfo {volume} {83}},\ \bibinfo {pages} {075125} (\bibinfo {year}
  {2011})%
  \bibAnnoteFile{NoStop}{LNO-optic-Stewart}%
\bibitem{LNO-optic-Stewart-Arxiv}%
  \BibitemOpen
  \bibfield{author}{%
  \bibinfo {author} {\bibfnamefont{M.~K.}\ \bibnamefont{Stewart}}, \bibinfo
  {author} {\bibfnamefont{J.}~\bibnamefont{Liu}}, \bibinfo {author}
  {\bibfnamefont{R.~K.}\ \bibnamefont{Smith}}, \bibinfo {author}
  {\bibfnamefont{B.~C.}\ \bibnamefont{Chapler}}, \bibinfo {author}
  {\bibfnamefont{C.~H.}\ \bibnamefont{Yee}}, \bibinfo {author}
  {\bibfnamefont{K.}~\bibnamefont{Haule}}, \bibinfo {author}
  {\bibfnamefont{J.}~\bibnamefont{Chakhalian}},\ and\ \bibinfo {author}
  {\bibfnamefont{D.~N.}\ \bibnamefont{Basov}},\ }%
  \Eprint{http://arxiv.org/abs/1005.3314}{arXiv:1005.3314}%
  \bibAnnoteFile{NoStop}{LNO-optic-Stewart-Arxiv}%
\bibitem{LDA+DMFT-Aichhorn}%
  \BibitemOpen
  \bibfield{author}{%
  \bibinfo {author} {\bibfnamefont{M.}~\bibnamefont{Aichhorn}}, \bibinfo
  {author} {\bibfnamefont{L.}~\bibnamefont{Pourovskii}}, \bibinfo {author}
  {\bibfnamefont{V.}~\bibnamefont{Vildosola}}, \bibinfo {author}
  {\bibfnamefont{M.}~\bibnamefont{Ferrero}}, \bibinfo {author}
  {\bibfnamefont{O.}~\bibnamefont{Parcollet}}, \bibinfo {author}
  {\bibfnamefont{T.}~\bibnamefont{Miyake}}, \bibinfo {author}
  {\bibfnamefont{A.}~\bibnamefont{Georges}},\ and\ \bibinfo {author}
  {\bibfnamefont{S.}~\bibnamefont{Biermann}},\ }%
  \bibfield{journal}{%
  \Doi{10.1103/PhysRevB.80.085101}{\bibinfo {journal} {Phys. Rev. B}}\ }%
  \textbf{\bibinfo {volume} {80}},\ \bibinfo {pages} {085101} (\bibinfo {year}
  {2009})%
  \bibAnnoteFile{NoStop}{LDA+DMFT-Aichhorn}%
\bibitem{LaFeAsO-Markus}%
  \BibitemOpen
  \bibfield{author}{%
  \bibinfo {author} {\bibfnamefont{M.}~\bibnamefont{Aichhorn}}, \bibinfo
  {author} {\bibfnamefont{L.}~\bibnamefont{Pourovskii}},\ and\ \bibinfo
  {author} {\bibfnamefont{A.}~\bibnamefont{Georges}},\ }%
  \bibinfo {note} {to appear in Phys. Rev. B},\
  \Eprint{http://arxiv.org/abs/1104.4361}{arXiv:1104.4361}%
  \bibAnnoteFile{NoStop}{LaFeAsO-Markus}%
\bibitem{UJ-Terakura}%
  \BibitemOpen
  \bibfield{author}{%
  \bibinfo {author} {\bibfnamefont{I.}~\bibnamefont{Solovyev}}, \bibinfo
  {author} {\bibfnamefont{N.}~\bibnamefont{Hamada}},\ and\ \bibinfo {author}
  {\bibfnamefont{K.}~\bibnamefont{Terakura}},\ }%
  \bibfield{journal}{%
  \Doi{10.1103/PhysRevB.53.7158}{\bibinfo {journal} {Phys. Rev. B}}\ }%
  \textbf{\bibinfo {volume} {53}},\ \bibinfo {pages} {7158} (\bibinfo {year}
  {1996})%
  \bibAnnoteFile{NoStop}{UJ-Terakura}%
\bibitem{LNO-XPS-Horiba}%
  \BibitemOpen
  \bibfield{author}{%
  \bibinfo {author} {\bibfnamefont{K.}~\bibnamefont{Horiba}}, \bibinfo {author}
  {\bibfnamefont{R.}~\bibnamefont{Eguchi}}, \bibinfo {author}
  {\bibfnamefont{M.}~\bibnamefont{Taguchi}}, \bibinfo {author}
  {\bibfnamefont{A.}~\bibnamefont{Chainani}}, \bibinfo {author}
  {\bibfnamefont{A.}~\bibnamefont{Kikkawa}}, \bibinfo {author}
  {\bibfnamefont{Y.}~\bibnamefont{Senba}}, \bibinfo {author}
  {\bibfnamefont{H.}~\bibnamefont{Ohashi}},\ and\ \bibinfo {author}
  {\bibfnamefont{S.}~\bibnamefont{Shin}},\ }%
  \bibfield{journal}{%
  \Doi{10.1103/PhysRevB.76.155104}{\bibinfo {journal} {Phys. Rev. B}}\ }%
  \textbf{\bibinfo {volume} {76}},\ \bibinfo {pages} {155104} (\bibinfo {year}
  {2007})%
  \bibAnnoteFile{NoStop}{LNO-XPS-Horiba}%
\bibitem{TRIQS}%
  \BibitemOpen
  \bibfield{author}{%
  \bibinfo {author} {\bibfnamefont{M.}~\bibnamefont{Ferrero}}\ and\ \bibinfo
  {author} {\bibfnamefont{O.}~\bibnamefont{Parcollet}},\ }%
  \enquote{\bibinfo {title} {{TRIQS}: a {T}oolkit for {R}esearch in
  {I}nteracting {Q}uantum {S}ystems},}\ \url{http://ipht.cea.fr/triqs}%
  \bibAnnoteFile{NoStop}{TRIQS}%
\bibitem{CTQMC-Werner}%
  \BibitemOpen
  \bibfield{author}{%
  \bibinfo {author} {\bibfnamefont{P.}~\bibnamefont{Werner}}, \bibinfo {author}
  {\bibfnamefont{A.}~\bibnamefont{Comanac}}, \bibinfo {author}
  {\bibfnamefont{L.}~\bibnamefont{de' Medici}}, \bibinfo {author}
  {\bibfnamefont{M.}~\bibnamefont{Troyer}},\ and\ \bibinfo {author}
  {\bibfnamefont{A.~J.}\ \bibnamefont{Millis}},\ }%
  \bibfield{journal}{%
  \Doi{10.1103/PhysRevLett.97.076405}{\bibinfo {journal} {Phys. Rev. Lett.}}\
  }%
  \textbf{\bibinfo {volume} {97}},\ \bibinfo {pages} {076405} (\bibinfo {year}
  {2006})%
  \bibAnnoteFile{NoStop}{CTQMC-Werner}%
\bibitem{CTQMC-Gull-review}%
  \BibitemOpen
  \bibfield{author}{%
  \bibinfo {author} {\bibfnamefont{E.}~\bibnamefont{Gull}}, \bibinfo {author}
  {\bibfnamefont{A.~J.}\ \bibnamefont{Millis}}, \bibinfo {author}
  {\bibfnamefont{A.~I.}\ \bibnamefont{Lichtenstein}}, \bibinfo {author}
  {\bibfnamefont{A.~N.}\ \bibnamefont{Rubtsov}}, \bibinfo {author}
  {\bibfnamefont{M.}~\bibnamefont{Troyer}},\ and\ \bibinfo {author}
  {\bibfnamefont{P.}~\bibnamefont{Werner}},\ }%
  \bibfield{journal}{%
  \Doi{10.1103/RevModPhys.83.349}{\bibinfo {journal} {Rev. Mod. Phys.}}\ }%
  \textbf{\bibinfo {volume} {83}},\ \bibinfo {pages} {349} (\bibinfo {year}
  {2011})%
  \bibAnnoteFile{NoStop}{CTQMC-Gull-review}%
\bibitem{Legendre}%
  \BibitemOpen
  \bibfield{author}{%
  \bibinfo {author} {\bibfnamefont{L.}~\bibnamefont{Boehnke}}, \bibinfo
  {author} {\bibfnamefont{H.}~\bibnamefont{Hafermann}}, \bibinfo {author}
  {\bibfnamefont{M.}~\bibnamefont{Ferrero}}, \bibinfo {author}
  {\bibfnamefont{F.}~\bibnamefont{Lechermann}},\ and\ \bibinfo {author}
  {\bibfnamefont{O.}~\bibnamefont{Parcollet}},\ }%
  \bibinfo {note} {to appear in Phys. Rev. B},\
  \Eprint{http://arxiv.org/abs/1104.3215}{arXiv:1104.3215}%
  \bibAnnoteFile{NoStop}{Legendre}%
\bibitem{Beach}%
  \BibitemOpen
  \bibfield{author}{%
  \bibinfo {author} {\bibfnamefont{K.~S.~D.}\ \bibnamefont{Beach}},\ }%
  \Eprint{http://arxiv.org/abs/cond-mat/0403055}{arXiv:cond-mat/0403055}%
  \bibAnnoteFile{NoStop}{Beach}%
\bibitem{Pade}%
  \BibitemOpen
  \bibfield{author}{%
  \bibinfo {author} {\bibfnamefont{H.~J.}\ \bibnamefont{Vidberg}}\ and\
  \bibinfo {author} {\bibfnamefont{J.~W.}\ \bibnamefont{Serene}},\ }%
  \bibfield{journal}{%
  \bibinfo {journal} {Journal of Low Temperature Physics}\ }%
  \textbf{\bibinfo {volume} {29}},\ \bibinfo {pages} {179} (\bibinfo {year}
  {1977})%
  \bibAnnoteFile{NoStop}{Pade}%
\bibitem{byczuk_kinks_2007}%
  \BibitemOpen
  \bibfield{author}{%
  \bibinfo {author} {\bibfnamefont{K.}~\bibnamefont{Byczuk}}, \bibinfo {author}
  {\bibfnamefont{M.}~\bibnamefont{Kollar}}, \bibinfo {author}
  {\bibfnamefont{K.}~\bibnamefont{Held}}, \bibinfo {author}
  {\bibfnamefont{Y.}~\bibnamefont{Yang}}, \bibinfo {author}
  {\bibfnamefont{I.~A.}\ \bibnamefont{Nekrasov}}, \bibinfo {author}
  {\bibfnamefont{T.}~\bibnamefont{Pruschke}},\ and\ \bibinfo {author}
  {\bibfnamefont{D.}~\bibnamefont{Vollhardt}},\ }%
  \bibfield{journal}{%
  \Doi{10.1038/nphys538}{\bibinfo {journal} {Nat. Phys.}}\ }%
  \textbf{\bibinfo {volume} {3}},\ \bibinfo {pages} {168} (\bibinfo {year}
  {2007})%
  \bibAnnoteFile{NoStop}{byczuk_kinks_2007}%
\bibitem{SVO-LDADMFT-Nekrasov}%
  \BibitemOpen
  \bibfield{author}{%
  \bibinfo {author} {\bibfnamefont{I.~A.}\ \bibnamefont{Nekrasov}}, \bibinfo
  {author} {\bibfnamefont{K.}~\bibnamefont{Held}}, \bibinfo {author}
  {\bibfnamefont{G.}~\bibnamefont{Keller}}, \bibinfo {author}
  {\bibfnamefont{D.~E.}\ \bibnamefont{Kondakov}}, \bibinfo {author}
  {\bibfnamefont{T.}~\bibnamefont{Pruschke}}, \bibinfo {author}
  {\bibfnamefont{M.}~\bibnamefont{Kollar}}, \bibinfo {author}
  {\bibfnamefont{O.~K.}\ \bibnamefont{Andersen}}, \bibinfo {author}
  {\bibfnamefont{V.~I.}\ \bibnamefont{Anisimov}},\ and\ \bibinfo {author}
  {\bibfnamefont{D.}~\bibnamefont{Vollhardt}},\ }%
  \bibfield{journal}{%
  \Doi{10.1103/PhysRevB.73.155112}{\bibinfo {journal} {Phys. Rev. B}}\ }%
  \textbf{\bibinfo {volume} {73}},\ \bibinfo {pages} {155112} (\bibinfo {year}
  {2006})%
  \bibAnnoteFile{NoStop}{SVO-LDADMFT-Nekrasov}%
\bibitem{uhrig_kink_prl_2009}%
  \BibitemOpen
  \bibfield{author}{%
  \bibinfo {author} {\bibfnamefont{C.}~\bibnamefont{Raas}}, \bibinfo {author}
  {\bibfnamefont{P.}~\bibnamefont{Grete}},\ and\ \bibinfo {author}
  {\bibfnamefont{G.~S.}\ \bibnamefont{Uhrig}},\ }%
  \bibfield{journal}{%
  \Doi{10.1103/PhysRevLett.102.076406}{\bibinfo {journal} {Phys. Rev. Lett.}}\
  }%
  \textbf{\bibinfo {volume} {102}},\ \bibinfo {pages} {076406} (\bibinfo {year}
  {2009})%
  \bibAnnoteFile{NoStop}{uhrig_kink_prl_2009}%
\bibitem{Model-Poteryaev}%
  \BibitemOpen
  \bibfield{author}{%
  \bibinfo {author} {\bibfnamefont{A.~I.}\ \bibnamefont{Poteryaev}}, \bibinfo
  {author} {\bibfnamefont{M.}~\bibnamefont{Ferrero}}, \bibinfo {author}
  {\bibfnamefont{A.}~\bibnamefont{Georges}},\ and\ \bibinfo {author}
  {\bibfnamefont{O.}~\bibnamefont{Parcollet}},\ }%
  \bibfield{journal}{%
  \Doi{10.1103/PhysRevB.78.045115}{\bibinfo {journal} {Phys. Rev. B}}\ }%
  \textbf{\bibinfo {volume} {78}},\ \bibinfo {pages} {045115} (\bibinfo {year}
  {2008})%
  \bibAnnoteFile{NoStop}{Model-Poteryaev}%
\bibitem{balents_MIT_arXiv}%
  \BibitemOpen
  \bibfield{author}{%
  \bibinfo {author} {\bibfnamefont{S.}~\bibnamefont{{Lee}}}, \bibinfo {author}
  {\bibfnamefont{R.}~\bibnamefont{{Chen}}},\ and\ \bibinfo {author}
  {\bibfnamefont{L.}~\bibnamefont{{Balents}}},\ }%
  \Eprint{http://arxiv.org/abs/1107.0724}{arXiv:1107.0724}%
  \bibAnnoteFile{NoStop}{balents_MIT_arXiv}%
\bibitem{hasan_prl_2004}%
  \BibitemOpen
  \bibfield{author}{%
  \bibinfo {author} {\bibfnamefont{M.~Z.}\ \bibnamefont{Hasan}}, \bibinfo
  {author} {\bibfnamefont{Y.-D.}\ \bibnamefont{Chuang}}, \bibinfo {author}
  {\bibfnamefont{D.}~\bibnamefont{Qian}}, \bibinfo {author}
  {\bibfnamefont{Y.~W.}\ \bibnamefont{Li}}, \bibinfo {author}
  {\bibfnamefont{Y.}~\bibnamefont{Kong}}, \bibinfo {author}
  {\bibfnamefont{A.}~\bibnamefont{Kuprin}}, \bibinfo {author}
  {\bibfnamefont{A.~V.}\ \bibnamefont{Fedorov}}, \bibinfo {author}
  {\bibfnamefont{R.}~\bibnamefont{Kimmerling}}, \bibinfo {author}
  {\bibfnamefont{E.}~\bibnamefont{Rotenberg}}, \bibinfo {author}
  {\bibfnamefont{K.}~\bibnamefont{Rossnagel}}, \bibinfo {author}
  {\bibfnamefont{Z.}~\bibnamefont{Hussain}}, \bibinfo {author}
  {\bibfnamefont{H.}~\bibnamefont{Koh}}, \bibinfo {author}
  {\bibfnamefont{N.~S.}\ \bibnamefont{Rogado}}, \bibinfo {author}
  {\bibfnamefont{M.~L.}\ \bibnamefont{Foo}},\ and\ \bibinfo {author}
  {\bibfnamefont{R.~J.}\ \bibnamefont{Cava}},\ }%
  \bibfield{journal}{%
  \Doi{10.1103/PhysRevLett.92.246402}{\bibinfo {journal} {Phys. Rev. Lett.}}\
  }%
  \textbf{\bibinfo {volume} {92}},\ \bibinfo {pages} {246402} (\bibinfo {year}
  {2004})%
  \bibAnnoteFile{NoStop}{hasan_prl_2004}%
\bibitem{yang_prl_2005}%
  \BibitemOpen
  \bibfield{author}{%
  \bibinfo {author} {\bibfnamefont{H.-B.}\ \bibnamefont{Yang}}, \bibinfo
  {author} {\bibfnamefont{Z.-H.}\ \bibnamefont{Pan}}, \bibinfo {author}
  {\bibfnamefont{A.~K.~P.}\ \bibnamefont{Sekharan}}, \bibinfo {author}
  {\bibfnamefont{T.}~\bibnamefont{Sato}}, \bibinfo {author}
  {\bibfnamefont{S.}~\bibnamefont{Souma}}, \bibinfo {author}
  {\bibfnamefont{T.}~\bibnamefont{Takahashi}}, \bibinfo {author}
  {\bibfnamefont{R.}~\bibnamefont{Jin}}, \bibinfo {author}
  {\bibfnamefont{B.~C.}\ \bibnamefont{Sales}}, \bibinfo {author}
  {\bibfnamefont{D.}~\bibnamefont{Mandrus}}, \bibinfo {author}
  {\bibfnamefont{A.~V.}\ \bibnamefont{Fedorov}}, \bibinfo {author}
  {\bibfnamefont{Z.}~\bibnamefont{Wang}},\ and\ \bibinfo {author}
  {\bibfnamefont{H.}~\bibnamefont{Ding}},\ }%
  \bibfield{journal}{%
  \Doi{10.1103/PhysRevLett.95.146401}{\bibinfo {journal} {Phys. Rev. Lett.}}\
  }%
  \textbf{\bibinfo {volume} {95}},\ \bibinfo {pages} {146401} (\bibinfo {year}
  {2005})%
  \bibAnnoteFile{NoStop}{yang_prl_2005}%
\bibitem{nicolaou_prl_2010}%
  \BibitemOpen
  \bibfield{author}{%
  \bibinfo {author} {\bibfnamefont{A.}~\bibnamefont{Nicolaou}}, \bibinfo
  {author} {\bibfnamefont{V.}~\bibnamefont{Brouet}}, \bibinfo {author}
  {\bibfnamefont{M.}~\bibnamefont{Zacchigna}}, \bibinfo {author}
  {\bibfnamefont{I.}~\bibnamefont{Vobornik}}, \bibinfo {author}
  {\bibfnamefont{A.}~\bibnamefont{Tejeda}}, \bibinfo {author}
  {\bibfnamefont{A.}~\bibnamefont{Taleb-Ibrahimi}}, \bibinfo {author}
  {\bibfnamefont{P.}~\bibnamefont{Le~F\`evre}}, \bibinfo {author}
  {\bibfnamefont{F.}~\bibnamefont{Bertran}}, \bibinfo {author}
  {\bibfnamefont{S.}~\bibnamefont{H\'ebert}}, \bibinfo {author}
  {\bibfnamefont{H.}~\bibnamefont{Muguerra}},\ and\ \bibinfo {author}
  {\bibfnamefont{D.}~\bibnamefont{Grebille}},\ }%
  \bibfield{journal}{%
  \Doi{10.1103/PhysRevLett.104.056403}{\bibinfo {journal} {Phys. Rev. Lett.}}\
  }%
  \textbf{\bibinfo {volume} {104}},\ \bibinfo {pages} {056403} (\bibinfo {year}
  {2010})%
  \bibAnnoteFile{NoStop}{nicolaou_prl_2010}%
\bibitem{ALPS}%
  \BibitemOpen
  \bibfield{author}{%
  \bibinfo {author} {\bibfnamefont{B.}~\bibnamefont{Bauer}}, \bibinfo {author}
  {\bibfnamefont{L.~D.}\ \bibnamefont{Carr}}, \bibinfo {author}
  {\bibfnamefont{H.~G.}\ \bibnamefont{Evertz}}, \bibinfo {author}
  {\bibfnamefont{A.}~\bibnamefont{Feiguin}}, \bibinfo {author}
  {\bibfnamefont{J.}~\bibnamefont{Freire}}, \bibinfo {author}
  {\bibfnamefont{S.}~\bibnamefont{Fuchs}}, \bibinfo {author}
  {\bibfnamefont{L.}~\bibnamefont{Gamper}}, \bibinfo {author}
  {\bibfnamefont{J.}~\bibnamefont{Gukelberger}}, \bibinfo {author}
  {\bibfnamefont{E.}~\bibnamefont{Gull}}, \bibinfo {author}
  {\bibfnamefont{S.}~\bibnamefont{Guertler}}, \bibinfo {author}
  {\bibfnamefont{A.}~\bibnamefont{Hehn}}, \bibinfo {author}
  {\bibfnamefont{R.}~\bibnamefont{Igarashi}}, \bibinfo {author}
  {\bibfnamefont{S.~V.}\ \bibnamefont{Isakov}}, \bibinfo {author}
  {\bibfnamefont{D.}~\bibnamefont{Koop}}, \bibinfo {author}
  {\bibfnamefont{P.~N.}\ \bibnamefont{Ma}}, \bibinfo {author}
  {\bibfnamefont{P.}~\bibnamefont{Mates}}, \bibinfo {author}
  {\bibfnamefont{H.}~\bibnamefont{Matsuo}}, \bibinfo {author}
  {\bibfnamefont{O.}~\bibnamefont{Parcollet}}, \bibinfo {author}
  {\bibfnamefont{G.}~\bibnamefont{Paw{\l}owski}}, \bibinfo {author}
  {\bibfnamefont{J.~D.}\ \bibnamefont{Picon}}, \bibinfo {author}
  {\bibfnamefont{L.}~\bibnamefont{Pollet}}, \bibinfo {author}
  {\bibfnamefont{E.}~\bibnamefont{Santos}}, \bibinfo {author}
  {\bibfnamefont{V.~W.}\ \bibnamefont{Scarola}}, \bibinfo {author}
  {\bibfnamefont{U.}~\bibnamefont{Schollw{\"o}ck}}, \bibinfo {author}
  {\bibfnamefont{C.}~\bibnamefont{Silva}}, \bibinfo {author}
  {\bibfnamefont{B.}~\bibnamefont{Surer}}, \bibinfo {author}
  {\bibfnamefont{S.}~\bibnamefont{Todo}}, \bibinfo {author}
  {\bibfnamefont{S.}~\bibnamefont{Trebst}}, \bibinfo {author}
  {\bibfnamefont{M.}~\bibnamefont{Troyer}}, \bibinfo {author}
  {\bibfnamefont{M.~L.}\ \bibnamefont{Wall}}, \bibinfo {author}
  {\bibfnamefont{P.}~\bibnamefont{Werner}},\ and\ \bibinfo {author}
  {\bibfnamefont{S.}~\bibnamefont{Wessel}},\ }%
  \bibfield{journal}{%
  \Doi{10.1088/1742-5468/2011/05/P05001}{\bibinfo {journal} {J. Stat. Mech.:
  Theor. Exp.}}\ }%
  \textbf{\bibinfo {volume} {2011}},\ \bibinfo {pages} {P05001} (\bibinfo
  {year} {2011})%
  \bibAnnoteFile{NoStop}{ALPS}%
\end{thebibliography}
\end{document}